\documentclass[11pt,a4paper]{amsart}
\usepackage{amsmath}
\usepackage{amssymb}
\usepackage{amsfonts}
\usepackage[toc,page]{appendix}
\usepackage{hyperref}

\def\={\ =\ }

\newcommand{\be}{\begin{equation}}
\newcommand{\ee}{\end{equation}}
\newcommand{\beq}{\begin{equation}}
\newcommand{\eeq}{\end{equation}}
\newcommand{\bea}{\begin{eqnarray}}
\newcommand{\eea}{\end{eqnarray}}

\def\ba{\begin{eqnarray}}
\def\ea{\end{eqnarray}}

\theoremstyle{plain}

\newtheorem{proposition*}{Proposition}
\newtheorem*{remark}{Remark}

\numberwithin{equation}{section}

\input{tcilatex}

\begin{document}
\title[]{Exact solution of Chern-Simons-matter matrix models with
characteristic/orthogonal polynomials}
\author{Miguel Tierz}
\address{Departamento de Matem\'{a}tica, Grupo de F\'{\i}sica Matem\'{a}%
tica, Faculdade de Ci\^{e}ncias, Universidade de Lisboa, Campo Grande, Edif%
\'{\i}cio C6, 1749-016 Lisboa, Portugal.}
\email{tierz@fc.ul.pt}
\maketitle

\begin{abstract}
We solve for finite $N$ the matrix model of 
supersymmetric $U(N)$ Chern-Simons theory coupled to $N_{f}$ fundamental and 
$N_{f}$ anti-fundamental chiral multiplets of $R$-charge $1/2$ and of mass $%
m $, by identifying it with an average of inverse characteristic polynomials
in a Stieltjes-Wigert ensemble. This requires the computation of the Cauchy
transform of the Stieltjes-Wigert polynomials, which we carry out, finding a
relationship with Mordell integrals, and hence with previous analytical
results on the matrix model. The semiclassical limit of the model is
expressed, for arbitrary $N_{f},$ in terms of a single Hermite polynomial.
This result also holds for more general matter content, involving matrix
models with double-sine functions.
\end{abstract}

\section{Introduction}

The study of supersymmetric gauge theories has been rekindled in recent
years with considerable progress mainly due to two reasons: on one hand, the
development of localization of supersymmetric gauge theories \cite%
{Pestun:2007rz} (\cite{Pestun:2014mja,Hosomichi:2015jta} for recent reviews)
and, on the other hand, the availability of powerful analytical tools to
explicitly handle matrix models, a long-standing area of mathematical and
theoretical physics \cite{For}. These two subjects are now intimately
related since the localization method manages to reduce the original
functional integral describing a quantum field theory into a much simpler
matrix integral. Thus, it enormously reduces the task of computing
observables in a supersymmetric gauge theory, but there still remains the
issue of explicitly computing $N$ integrations, which is typically not an
straightforward task and requires the use of specific matrix model
techniques.

The theory we shall focus on is ${\mathcal{N}}=2$ supersymmetric $U(N)$
Chern-Simons (CS) on the three-sphere $\mathbb{S}^{3}$ with $N_{f}$
fundamental and $N_{f}$ antifundamental chiral multiplets of mass $m$.
Indeed the partition function on $\mathbb{S}^{3}$ can be determined by the
localization techniques of Pestun, which were adapted to the 3d case in \cite%
{Kapustin:2009kz,Kapustin:2010xq,Jafferis:2010un,Hama:2010av}. In the case
of the partition function for $U(N)$ $\mathcal{N}=3$ Chern-Simons theory at
level $k$ coupled to $N_{f}$ fundamental and $\bar{N}_{f}$ anti-fundamental
chiral multiplets of $R$-charge $q$ the matrix model is \cite{Hama:2010av}%
\footnote{%
Notice that we have changed the sign of the Chern-Simons level with respect
to that in \cite{Hama:2010av} in order to make contact with our conventions.}%
\begin{equation}
Z=\frac{1}{N!}\int d^{N}\sigma \;\prod_{j=1}^{N}e^{i\pi k\sigma
_{j}^{2}}\left( s_{b=1}(i-iq-\sigma _{j})\right) ^{N_{f}}\left(
s_{b=1}(i-iq+\sigma _{j})\right) ^{\bar{N}_{f}}\prod_{i<j}^{N}(2\sinh \pi
(\sigma _{i}-\sigma _{j}))^{2},  \label{Zds}
\end{equation}%
where $s_{b=1}(\sigma )$ denotes the double sine function \cite%
{Jafferis:2010un,Hama:2010av} (and references therein). This matrix model
corresponds to the case where the matter chiral multiplets have $R$-charge $%
q $ and belong to the representation $R$ of the gauge group. The fact that
for $\mathcal{N}=3$ theories the R-symmetry is non-abelian allows us to fix
an R-charge which is not altered under the RG flow\footnote{%
Our formalism will implicitly allow for imaginary masses and thus one can
view the matrix model as corresponding to a $\mathcal{N}=2$ theory, for
which the $\mathcal{Z}$-extremization process \cite{Jafferis:2010un} has not
been carried out.}. In this paper, we focus on a comprehensive analyzing of
the case where $q=1/2$ and $R=r\oplus \overline{r}$, in which case, due to a
basic property of the double sine function, the matter contribution
simplifies in the following way \cite{Hama:2010av}%
\begin{equation}
\prod_{\rho \in r}s_{b=1}(\tfrac{i}{2}-\rho _{i}\hat{\sigma}_{i})\cdot
s_{b=1}(\tfrac{i}{2}+\rho _{i}\hat{\sigma}_{i})~=~\prod_{\rho \in r}\frac{1}{%
2\cosh \pi \rho _{i}\hat{\sigma}_{i}}\,,  \label{dsid}
\end{equation}%
which renders the matrix model equal to%
\begin{equation}
Z_{N_{f}}^{U(N)}=\frac{1}{\left( 2\pi \right) ^{N}N!}\int {d^{N}\!\mu }\frac{%
\prod_{i<j}4\sinh ^{2}(\frac{1}{2}(\mu _{i}-\mu _{j}))\ e^{-\frac{1}{2g}%
\sum_{i}\mu _{i}^{2}+i\eta \sum_{i}\mu _{i}}}{\prod_{i}\left( 2\cosh (\frac{1%
}{2}(\mu _{i}+m))\right) ^{N_{f}}}\ ,  \label{Z}
\end{equation}%
where $g=\frac{2\pi i}{k}$ with $k\in \mathbb{Z}$ the Chern-Simons level and 
$\mu _{i}/2\pi $ represent the eigenvalues of the scalar field $\sigma $
belonging to the three dimensional 
vector multiplet. In \eqref{Z} the radius $R$ of the three-sphere has been
set to one. It can be restored by rescaling $m\rightarrow mR$, $\mu
_{i}\rightarrow \mu _{i}R$. The partition function is periodic in imaginary
shifts of the mass, $Z(m+i2\pi n)=Z(m)$, for integer $n$. The addition of a
a Fayet-Iliopoulos term (FI) in the Lagrangian adds a linear term in the
potential of the matrix model \cite%
{Kapustin:2009kz,Kapustin:2010xq,Hama:2010av}. Thus $\eta $ is a real
parameter denoting our FI parameter. Notice that the variables in (\ref{Z})
are rescaled with a $2\pi $ factor with regards to those in \cite%
{Kapustin:2009kz,Kapustin:2010xq,Hama:2010av} and with regard to the ones in
(\ref{Zds}). That is, $\mu _{i}=2\pi \sigma _{i}$.

A variant of the matrix model (\ref{Z}), with $2N_{f\text{ }}$
hypermultiplets has been previously analyzed in \cite{BR,Russo:2014bda}. In 
\cite{Russo:2014bda}, the approach is to express (\ref{Z}) as a Hankel
determinant whose entries are (combinations of) Mordell integrals, which are
integrals of the type \cite{Mordell}%
\begin{equation}
I(l,m)=\int_{-\infty }^{\infty }d\mu \frac{e^{(l+1)\mu +m}}{1+e^{\mu +m}}%
e^{-\mu ^{2}/2g}.  \label{imor}
\end{equation}%
This integral $I$ (\ref{imor}) was computed by Mordell \cite{Mordell} for
general parameters. Typically, it is given in terms of infinite sums of the
theta-function type. However, in specific cases it assumes the form of a
Gauss's finite sum \cite{Mordell,Russo:2014bda}. These specific cases
precisely contain the one which is physically relevant: $g=2\pi i/k$ with $%
k\in \mathbb{Z}$. This works very well for $N_{f}=1$ and has been more
recently extended to higher flavour in \cite{GM} by studying parametric
derivatives of Mordell integrals. A large number of analytic results and
explicit tests of Giveon-Kutasov duality can be obtained in that way.

An alternative analytical approach is put forward in this paper. In
particular, instead of using the Mordell integral \cite{Mordell} method
developed in \cite{Russo:2014bda,GM}, we shall employ here the powerful
formalism of characteristic polynomials in random matrix ensembles, which
has been considerably developed over the last decade \cite%
{Fyodorov:2002ub,Fyodorov:2002jw}.

The paper is organized as follows. In the next Section we introduce the
necessary background on random matrices and averages of characteristic
polynomials on random matrix ensembles, together with its connections with
orthogonal polynomials. We shall then show, in Section 3, that the matrix
model (\ref{Z}) above can be solved in terms of such averages in a
Stieltjes-Wigert ensemble. Determinantal formulas for the model in \cite%
{Russo:2014bda} and for the model (\ref{Z}) with different masses are given
as well.

In Section 4, we characterize analytically the Stieltjes transform of the
Stieltjes-Wigert polynomials, which is the fundamental object in Section 2
and also the building block in the determinantal formulas that emerge in the 
$N_{f}>1$ case. The polynomials are obtained explicitly and shown to be
related to Mordell integrals (\ref{imor}). A number of particular cases for
both low and large $N$ are presented and consistency checks are carried out.
In particular, using modular properties of the Mordell integral \cite{Mock}
we also establish the relationship with the method of Hankel determinants of
Mordell integrals used in \cite{Russo:2014bda} and extended in \cite{GM}.

In Section 5, we study the semiclassical limit of the matrix model, finding
it to be given by a single Hermite polynomial. This holds for $N_{f}\geq 1$,
because the number of flavours appears in the spectral parameter of the
characteristic polynomial average (hence, in the variable of the Hermite
polynomial) and, therefore, no determinantal expression is required in this
limit for $N_{f}>1$. The same behavior is shown to hold also for the model (%
\ref{Zds}) and not just (\ref{Z}). Hence the solution of the model in the
semiclassical limit is still a single Hermite polynomial, only with a more
involved specialization of the variable, which becomes complex, of the
polynomial. We finally conclude with a summary and avenues for further
research.

\section{Random matrices and characteristic polynomials, definitions.}

\label{intro} 

In random matrix theory, ensembles of $N\times N$ Hermitian matrices $\{H\}$
are described by a measure $d\alpha $ with finite moments $\int_{\mathbb{R}%
}|x|^{k}d\alpha (x)<\infty $, $k=0,1,2,\cdots $, and the probability
distribution function for the eigenvalues $\{x_{i}=x_{i}(H)\}$ of matrices $%
H $ in the ensembles is of the form%
\begin{equation}
\text{\textrm{Prob}}_{\alpha ,N}(x)=\frac{1}{Z_{N}}\Delta (x)^{2}d\alpha (x)
\label{0.1}
\end{equation}%
where $d\alpha (x)=\prod_{i=1}^{N}d\alpha (x_{i})$, $\Delta (x)=\prod_{N\geq
i>j\geq 1}(x_{i}-x_{j})$ is the Vandermonde determinant for the $x_{i}$'s,
and $Z_{N}=\int \cdots \int \Delta (x)^{2}d\alpha (x)$ is the normalization
constant. Therefore the integrals that emerge from localization and after
the suitable change of variables, are precisely of the same form as $Z_{N}$.
Choosing the measure $d\alpha (x)$ to be of the form $d\alpha
(x)=e^{-V(x)}dx $ where $V(x)$ is known as the potential of the matrix
model, one can also write (\ref{0.1}) alternatively as 
\begin{equation}
\frac{1}{Z_{N}}\det \left( f_{i}(x_{j})\right) _{i,j=1}^{N}\det \left(
g_{i}(x_{j})\right) _{i,j=1}^{N}\ \prod_{j=1}^{N}dx_{j}  \label{jointPDF2}
\end{equation}%
where%
\begin{equation}
f_{i}(x)=x^{i-1},\qquad g_{i}(x)=x^{i-1}e^{-V(x)}.  \label{jointPDF2bis}
\end{equation}%
Indeed, this follows upon recognizing \eqref{0.1} to be basically a product
of two Vandermonde determinants. In the last decade or so, the mathematical
apparatus of random matrix theory has been developed considerably \cite{For}
and, in particular, there has been much progress in the study of averages of
products and ratios of the characteristic polynomials%
\begin{equation}
D_{N}[\mu ,H]=\prod_{i=1}^{N}(\mu -x_{i}(H)),  \label{D}
\end{equation}%
of random matrices with respect to a number of ensembles \cite%
{Fyodorov:2002ub,Fyodorov:2002jw}\footnote{%
The literature on characteristic polynomials in random matrix theory is much
larger and we only quote the works that specifically treat the cases which
we will show describe Chern-Simons-matter.}. The variable $\mu $ in (\ref{D}%
) is a spectral parameter, which will be further characterized below.

Notice now that, as happens in the case of pure Chern-Simons theory \cite%
{Tierz}, the change of variables%
\begin{equation}
z_{i}=ce^{\mu _{i}}\,,\hspace{2cm}c=e^{g\left( N+\frac{N_{f}}{2}-i\eta
\right) }\,,  \label{CV}
\end{equation}%
brings the matrix model (\ref{Z}) in standard random-matrix form \cite%
{Russo:2014bda,GM}%
\begin{equation}
Z_{N_{f}}^{U(N)}=\frac{e^{-\frac{gN}{2}(N+\frac{N_{f}}{2}-i\eta )(N-\frac{%
5N_{f}}{2}+i\eta )}e^{-3mNN_{f}/2}}{(2\pi )^{N}N!}\int_{\left[ 0,\infty
\right) ^{N}}{d^{N}z}\ \prod_{i<j}(z_{i}-z_{j})^{2}\frac{e^{-\frac{1}{2g}%
\sum_{i}(\ln z_{i})^{2}}}{\prod_{i}\left( ce^{-m}+z_{i}\right) ^{N_{f}}}.
\label{CSM}
\end{equation}%
We recall that the Stieltjes-Wigert ensemble is characterized by $V(z)=\frac{%
1}{2g}\ln ^{2}z$ and solves exactly pure $U(N)$ Chern-Simons theory on $%
\mathbb{S}^{3}$, as was first shown in \cite{Tierz}. One possible
interpretation of the model (\ref{CSM}) is as the normalization constant of
the random matrix ensemble with potential $V(z)=\frac{1}{2g}\ln
^{2}z+N_{f}\ln \left( 1+z\frac{e^{m}}{c}\right) $, where $z\in \left(
0,\infty \right) $. However, this logarithmic deformation of the
Stieltjes-Wigert weight does not have a known closed system of orthogonal
polynomials. Therefore, in principle a full analysis of (\ref{CSM}) can not
be carried out with the same level of detail as with the much simpler
Stieltjes-Wigert ensemble.

\begin{remark}
The main idea in this paper is that a more detailed analytical
characterization of the Chern-Simons-matter matrix model can be actually
obtained by interpreting (\ref{CSM}) as the average of the inverse of a
characteristic polynomial (to be defined in what follows) in a
Stieltjes-Wigert ensemble.
\end{remark}

Let us then begin by introducing the simplest and most basic result on
characteristic polynomials. Notice that one can associate the average
characteristic polynomial to the probability distribution \eqref{jointPDF2}--%
\eqref{jointPDF2bis}%
\begin{equation}
P_{n}(z)=\frac{1}{\tilde{Z}_{n}}\int_{-\infty }^{\infty }\ldots
\int_{-\infty }^{\infty }\left( \prod_{j=1}^{n}(z-x_{j})\right) \ \det
\left( f_{i}(x_{j})\right) _{i,j=1}^{n}\det \left( g_{i}(x_{j})\right)
_{i,j=1}^{n}\prod_{j=1}^{n}dx_{j}\text{ }.  \label{averagecharpol:intro}
\end{equation}%
It then follows from a classical calculation of Heine, see e.g.\ \cite{Deift}%
, that $P_{n}$ can be characterized as the $n$th monic orthogonal polynomial
with respect to the weight function $e^{-V(x)}$. This means that the
polynomials $P_{n}$ satisfy the conditions%
\begin{equation*}
P_{n}(x)=x^{n}+O(x^{n-1})
\end{equation*}%
for all $n\in 
\mathbb{N}
$ and%
\begin{equation}
\int_{-\infty }^{\infty }P_{n}(x)P_{m}(x)e^{-V(x)}\ dx=c_{n}c_{m}\delta
_{m,n}  \label{def:cn}
\end{equation}%
for all $n,m\in 
\mathbb{N}
$, for certain $c_{n}\in 
\mathbb{R}
$.

Likewise, one can define the average inverse characteristic polynomial
corresponding to the probability distribution \eqref{jointPDF2}--%
\eqref{jointPDF2bis} as \cite{Fyodorov:2002ub}%
\begin{equation}
Q_{n}(z)=\frac{1}{\tilde{Z}_{n}}\int_{-\infty }^{\infty }\ldots
\int_{-\infty }^{\infty }\left( \prod_{j=1}^{n}(z-x_{j})^{-1}\right) \ \det
\left( f_{i}(x_{j})\right) _{i,j=1}^{n}\det \left( g_{i}(x_{j})\right)
_{i,j=1}^{n}\prod_{j=1}^{n}dx_{j},  \label{averageinvcharpol:intro}
\end{equation}%
for $z\in 
\mathbb{C}
\setminus 
\mathbb{R}
$. Then it holds that%
\begin{equation}
Q_{n}(z)=\int_{-\infty }^{\infty }\frac{P_{n-1}(x)e^{-V(x)}}{z-x}\ dx.
\label{Cauchy}
\end{equation}%
Thus $Q_{n}(z)$ is the Stieltjes (or Cauchy) transform of the $P_{n-1}(x)$
polynomials multiplied by their weight function $e^{-V(x)}$. The normalizing
partition function in (\ref{averageinvcharpol:intro}) is, in terms of the
orthogonal polynomials $\tilde{Z}_{n}=N!\prod\nolimits_{k=0}^{N-1}c_{k}^{2}$
.

\section{Chern-Simons-matter models as correlation functions of
characteristic polynomials}

The first step is to notice that already (\ref{averageinvcharpol:intro}) can
be immediately identified with (\ref{CSM}) in the case $N_{f}=1$. More
precisely, we have that%
\begin{eqnarray}
\overline{Z}_{N_{f}=1}^{U(N)} &=&\frac{1}{\overline{Z}_{N_{f}=0}^{U(N)}}%
\int_{\left[ 0,\infty \right) ^{N}}{d^{N}z}\ \prod_{i<j}(z_{i}-z_{j})^{2}%
\frac{e^{-\frac{1}{2g}\sum_{i=1}^{N}(\ln z_{i})^{2}}}{\prod%
\nolimits_{i=1}^{N}\left( ce^{-m}+z_{i}\right) }  \notag \\
&=&\gamma _{N-1}\int_{0}^{\infty }\frac{\pi _{N-1}(x;q)e^{-\frac{1}{2g}(\ln
x)^{2}}}{\lambda +x}\ dx=:\gamma _{N-1}h_{N-1}(\lambda ),  \label{h-def}
\end{eqnarray}%
where $\pi _{N-1}(x)$ is the monic Stieltjes-Wigert polynomial \cite{Tierz},
a polynomial of order $N-1$, given explicitly in next Section, $\lambda
=ce^{-m}$ and $\gamma _{N}=-1/c_{N}^{2}$ and is a numerical prefactor that
will be appearing often below, in the determinantal expressions.

We emphasize that the polynomial $\gamma _{N-1}h_{N-1}(\lambda )$ computes
the integral in (\ref{CSM}), \textbf{normalized} by a matrix integral which
is the same ensemble but without the characteristic polynomial insertion.
That is, normalized by a Stieltjes-Wigert ensemble \cite{Tierz}%
\begin{equation}
\overline{Z}_{N_{f}=0}^{U(N)}=Z_{N}^{\mathrm{SW}}=\frac{1}{\left( 2\pi
g\right) ^{N/2}}\int_{\left[ 0,\infty \right) ^{N}}{d^{N}z}\
\prod_{i<j}(z_{i}-z_{j})^{2}e^{-\frac{1}{2g}\sum_{i=1}^{N}(\ln z_{i})^{2}}.
\label{ZSW0}
\end{equation}%
We will need its explicit expression below, so let us collect it here \cite[%
Eq. (2.13)]{Tierz}%
\begin{equation}
Z_{N}^{\mathrm{SW}}=N!q^{-\frac{N(2N-1)(2N+1)}{6}}\prod%
\limits_{j=1}^{N-1}(1-q^{j})^{N-j}.  \label{ZSW}
\end{equation}%
Thus, the polynomial $h_{N-1}(\lambda ),$ which is the Stieltjes (or Cauchy)
transform of the Stieltjes-Wigert polynomial $\pi _{N-1}(x;q)$%
\begin{equation}
h_{N-1}(\lambda ):=\int_{0}^{\infty }\frac{\pi _{N-1}(x;q)e^{-\frac{1}{2g}%
(\ln x)^{2}}dx}{\lambda +x},  \label{hn-1}
\end{equation}%
itself provides a solution, for all $N$, of the matrix integral for $N_{f}=1$%
, whereas with the Mordell integral method \cite{Russo:2014bda,GM} a $%
N\times N$ determinant has to be computed. To obtain (\ref{hn-1}) is
non-trivial and will be the subject of the next Section. Therefore, using
characteristic polynomials is essentially equivalent to having the
orthogonal polynomials for the Mordell ensemble \cite{Russo:2014bda} (\ref%
{CSM}).

\subsection{Determinantal formulas}

Suppose $1\leq N_{f}\leq N$ and let $\gamma _{n}=-1/c_{n}^{2}$, where $c_{n}$
is the normalization constant defined above (\ref{def:cn}). Then the
determinantal expression holds \cite{Fyodorov:2002ub,Fyodorov:2002jw}%
\begin{equation}
\left\langle \prod\limits_{j=1}^{N_{f}}D_{N}^{-1}\left[ \epsilon _{j},H%
\right] \right\rangle =\;\frac{(-1)^{\frac{N_{f}(N_{f}-1)}{2}%
}\prod_{j=N-N_{f}}^{N-1}\gamma _{j}}{\Delta (\epsilon _{1},...,\epsilon
_{N_{f}})}\left\vert 
\begin{array}{ccc}
h_{N-N_{f}}(\epsilon _{1}) & \ldots  & h_{N-1}(\epsilon _{1}) \\ 
\vdots  &  &  \\ 
h_{N-N_{f}}(\epsilon _{N_{f}}) & \ldots  & h_{N-1}(\epsilon _{N_{f}})%
\end{array}%
\right\vert ,  \label{averageOM}
\end{equation}%
where $D_{N}\left[ \epsilon _{j},H\right] $ is given by (\ref{D}), the
average is taken over a random matrix ensemble (\ref{0.1}) and $%
h_{n}(\epsilon )$ are the Stieltjes transform (\ref{Cauchy}) of the
polynomials orthogonal with regards to the measure $d\alpha (x)$ in (\ref%
{0.1}). In the case of the Stieltjes-Wigert ensemble, these are given by (%
\ref{h-def}). Normally, to avoid divergences, the spectral parameters $%
\epsilon _{j}$ are taken with a (small) imaginary part \cite{Fyodorov:2002ub}%
, as in the definition of the corresponding Stieltjes transform. The range
of integration in (\ref{Cauchy}) is $%
\mathbb{R}
^{+}$ in our case, since it is the domain of our random matrix ensemble. In
general, our parameter is either $\lambda >0$ or $\lambda \in 
\mathbb{C}
$,\footnote{%
There will be, for a given $N$ and $N_{f}$ at least one value of $k$ that
will imply $\lambda \in 
\mathbb{R}
$ and $\lambda <0,$ in which case one has to be careful understanding the
Cauchy transform as a principal value.} and, since $x\in 
\mathbb{R}
^{+}$, the corresponding Cauchy transform is well-defined.

Notice that this formalism is extensive enough to solve the more general
setting where all the masses of the multiplets are different. Namely, to%
\begin{equation}
\widehat{Z}_{N_{f}}^{U(N)}=\frac{1}{Z_{N}^{\mathrm{SW}}}\int_{\left[
0,\infty \right) ^{N}}{d^{N}z}\ \prod_{i<j}(z_{i}-z_{j})^{2}\frac{e^{-\frac{1%
}{2g}\sum_{i}(\ln z_{i})^{2}}}{\prod_{j=1}^{N_{f}}\prod_{i=1}^{N}\left(
ce^{-m_{j}}+z_{i}\right) }\ ,  \label{genZ}
\end{equation}%
for which, using (\ref{averageOM}), we find%
\begin{equation}
\widehat{Z}_{N_{f}}^{U(N)}=\frac{(-1)^{\frac{N_{f}(N_{f}-1)}{2}%
}\prod_{j=N-N_{f}}^{N-1}\gamma _{j}}{\Delta (\lambda _{1},...,\lambda
_{N_{f}})}\left\vert 
\begin{array}{ccc}
h_{N-N_{f}}(\lambda _{1}) & \ldots & h_{N-1}(\lambda _{1}) \\ 
\vdots &  &  \\ 
h_{N-N_{f}}(\lambda _{N_{f}}) & \ldots & h_{N-1}(\lambda _{N_{f}})%
\end{array}%
\right\vert ,  \label{genZdet}
\end{equation}%
where $\lambda _{j}=ce^{-m_{j}}$ and the polynomials are exactly (\ref{h-def}%
). Notice that this correlator admits an interpretation as $N_{f}$
non-compact branes at the different positions given by the $\lambda _{j}$ on
the Calabi-Yau threefold \textbf{P}$^{1}$ \cite{Hyun:2006dr}. For $N_{f}=1$
it can be interpreted as the amplitude of a non-compact brane.

The matrix model (\ref{CSM}) is a particular case of (\ref{genZ}) which
follows by taking $\lambda _{j}=ce^{-m}$ for $j=1,...,N_{f}.$ For this, as
usual with these determinants, we only need to consider the corresponding
limit in the determinantal expression (\ref{genZdet}), which can be easily
found by applying l'H\^{o}pital's rule \footnote{%
This is identical to the diagonal limit of the Christoffel-Darboux kernel of
a random matrix ensemble \cite{For}, and hence standard in random matrix
theory.}. In this way, we have that, for $N_{f}>1$ the determinant is the
Wronskian of the polynomials%
\begin{equation}
\overline{Z}_{N_{f}}^{U(N)}=(-1)^{\frac{N_{f}(N_{f}-1)}{2}%
}\prod_{j=N-N_{f}}^{N-1}\gamma _{j}\left\vert 
\begin{array}{ccc}
h_{N-1}(\lambda ) & \ldots & h_{N-N_{f}}(\lambda ) \\ 
h_{N-1}^{\prime }(\lambda ) &  & h_{N-N_{f}}^{\prime }(\lambda ) \\ 
\vdots &  &  \\ 
h_{N-1}^{(N_{f}-1)}(\lambda ) & \ldots & h_{N-N_{f}}^{(N_{f}-1)}(\lambda )%
\end{array}%
\right\vert .  \label{W}
\end{equation}%
Therefore, the characteristic polynomials method gives results for arbitrary 
$N$ and, for $N_{f}>1$ arbitrary, a $N_{f}\times N_{f}$ determinant is then
required.\ We emphasize that the difference with \cite{Russo:2014bda} is
that here the full $N$ dependence is encapsulated in the polynomial itself,
rather than in a determinant. \newline
We will characterize below the $\gamma _{j}$ prefactors for the
Stieltjes-Wigert polynomials but, given their definition as $\gamma
_{j}=-1/c_{j}^{2}$ and the inclusion of the partition function $%
Z_{N}=N!\prod\nolimits_{k=0}^{N-1}c_{k}^{2}$ , as a normalization factor in
the average of the characteristic polynomial, the simpler route will be to
directly cancel common factors.

\subsection{The theory with $2N_{f}$ Chiral Hypermultiplets}

This theory was analyzed in detail in \cite{Russo:2014bda} for $N_{f}=1$.
The partition function, expressed as a matrix integral, is%
\begin{equation}
Z_{2N_{f}}^{U(N)}=\frac{1}{\left( 2\pi \right) ^{N}N!}\int {d^{N}\!\mu }%
\frac{\prod_{i<j}4\sinh ^{2}(\frac{1}{2}(\mu _{i}-\mu _{j}))\ e^{-\frac{1}{2g%
}\sum_{i}\mu _{i}^{2}+i\eta \sum_{i}\mu _{i}}}{\prod_{i}\left( 4\cosh (\frac{%
1}{2}(\mu _{i}+m))\cosh (\frac{1}{2}(\mu _{i}-m))\right) ^{N_{f}}}\,,
\label{Zm2}
\end{equation}%
although in \cite{Russo:2014bda} we did not include the FI term ($\eta =0$).
The simple change of variables is \cite{BR,Russo:2014bda}%
\begin{equation}
z_{i}=ce^{\mu _{i}}\ ,\qquad c\equiv e^{g(N+N_{f}-i\eta )}\ ,
\label{changeofvariable2}
\end{equation}%
which recasts the integral in the form%
\begin{equation}
Z_{N_{f}}^{U(N)}=\frac{e^{-\frac{gN}{2}(N+N_{f}-i\eta )(N-5N_{f}+i\eta )}}{%
\left( 2\pi \right) ^{N}N!}\int_{\left[ 0,\infty \right) ^{N}}{d^{N}z}\
\prod_{i<j}(z_{i}-z_{j})^{2}\frac{e^{-\frac{1}{2g}\sum_{i}(\ln z_{i})^{2}}}{%
\prod_{i}\left( ce^{-m}+z_{i}\right) ^{N_{f}}\left( ce^{m}+z_{i}\right)
^{N_{f}}}\,.  \label{Z-2}
\end{equation}%
As shown in \cite{Russo:2014bda}, this matrix integral can be performed with
the help of Mordell integrals \cite{Mordell}. This method is also used in 
\cite{GM} for computations with $N_{f}>1$. Notice also that we have $2N_{f}$
and half of the masses are $m$ and the other half $-m$ (and hence, the
prefactors in (\ref{Z-2}) and (\ref{CSM}) are completely consistent).
Therefore, the solution in terms of orthogonal polynomials also follows from
(\ref{genZdet}) by considering two, instead of one, parameters $\lambda
=ce^{-m}$ and $\mu =ce^{m}$. We thus have, for $N_{f}=1$%
\begin{eqnarray}
\widetilde{Z}_{N_{f}=1}^{U(N)} &=&\frac{1}{Z_{N}^{\mathrm{SW}}}\int_{\left[
0,\infty \right) ^{N}}{d^{N}z}\ \prod_{i<j}(z_{i}-z_{j})^{2}\frac{e^{-\frac{1%
}{2g}\sum_{i}(\ln z_{i})^{2}}}{\prod_{i=1}^{N}\left( 1+z_{i}\frac{e^{m}}{c}%
\right) \left( 1+z_{i}\frac{e^{-m}}{c}\right) }  \label{Z2} \\
&=&\frac{-\gamma _{N-2}\gamma _{N-1}}{2c\sinh m}\left( h_{N-2}(\lambda
)h_{N-1}(\mu )-h_{N-1}(\lambda )h_{N-2}(\mu )\right) ,  \notag
\end{eqnarray}%
where again, as before, $\widetilde{Z}_{N_{f}=1}^{U(N)}$ refers to the
integral in (\ref{Z-2}) normalized by the SW partition function (\ref{ZSW0}%
). In the massless limit%
\begin{equation*}
\widetilde{Z}_{N_{f}=1}^{U(N)}(m=0)=-\gamma _{N-2}\gamma _{N-1}c^{-1}\left(
h_{N-2}(c)h_{N-1}^{\prime }(c)-h_{N-1}(c)h_{N-2}^{\prime }(c)\right) .
\end{equation*}%
We have thus established the solution of the matrix integrals in terms of
the polynomials $h_{n-1}(\lambda )$ (\ref{h-def}). Therefore, in the next
Section, we study and fully characterize analytically these polynomials,
which are Stieltjes transforms of the Stieltjes-Wigert polynomials.

\section{The Stieltjes/Cauchy transform of the Stieltjes-Wigert polynomials}

The Stieltjes--Wigert polynomials \cite{szego}-\cite{Wigert}%
\begin{equation}
P_{n}(x;q)=\frac{(-1)^{n}q^{n/2+1/4}}{\sqrt{(q;q)_{n}}}\sum_{j=0}^{n}\QATOPD[
] {n}{j}_{q}(-1)^{j}q^{j^{2}+j/2}x^{j},\quad n=0,1,\ldots  \label{SW}
\end{equation}%
are orthogonal with regards to the weight function%
\begin{equation}
\omega _{\mathrm{SW}}\left( x\right) =\frac{1}{\sqrt{\pi }}\overline{k}%
\mathrm{e}^{-\overline{k}^{2}\log ^{2}x},  \label{LN1}
\end{equation}%
with $q=e^{-1/2\overline{k}^{2}}$. We used the standard notation%
\begin{equation*}
(q;q)_{0}=1,\quad (q;q)_{n}=\prod_{j=1}^{n}(1-q^{j}),\quad n=1,2,\ldots
\end{equation*}%
and%
\begin{equation}
\QATOPD[ ] {n}{j}_{q}=\frac{(q;q)_{n}}{(q;q)_{j}(q;q)_{n-j}},\quad 0\leq
j\leq n.  \label{qbin}
\end{equation}%
There are other slightly different definitions of the polynomials, but (\ref%
{SW}) is the original by Wigert \cite{Wigert} and the most appropriate in
our context, since the corresponding weight function is (\ref{LN1}). \ Let
us first construct the monic version of the polynomials because, as we have
seen above, those are the ones that are identified with the average of
characteristic polynomials. To obtain the monic version of (\ref{SW}) we
need to divide it by $\widetilde{\gamma }_{n}=q^{n^{2}+n+1/4}/\sqrt{(q;q)_{n}%
}$ obtaining%
\begin{equation}
\pi _{n}\left( x\right) =\frac{P_{n}(x)}{\widetilde{\gamma }_{n}}%
=(-1)^{n}q^{-n^{2}-n/2}\sum_{j=0}^{n}\QATOPD[ ] {n}{j}%
_{q}(-1)^{j}q^{j^{2}+j/2}x^{j}.  \label{monicSW}
\end{equation}%
Notice that we will need the polynomials for $q$ roots of unity. While they
are better understood for $q<1$ (or $q>1$), they also hold for $q$ root of
unity \cite{AAK}. Notice for example, that the $q$-binomial coefficient is a
polynomial in $q$ and hence also well-defined \cite{AAK}. In general, the
whole four-parameter Askey-Wilson polynomials are studied also for $q$ root
of unity \cite{SZ} (the Stieltjes-Wigert are at the bottom of the Askey
classification scheme). In terms of basic hypergeometric functions they can
be written%
\begin{equation*}
\pi _{n}\left( x\right) =(-1)^{n}q^{-n^{2}-\frac{n}{2}}{_{1}\phi _{1}}%
(q^{-n},0;q,-q^{n+3/2}x).
\end{equation*}%
Because the polynomials (\ref{SW}) are actually orthonormal, it follows that
the $c_{k}$ coefficients in the orthogonality relationship (\ref{def:cn})
for the monic polynomials (\ref{monicSW}) are given such that $\gamma
_{n}=-1/c_{n}^{2}=-\widetilde{\gamma }_{n}^{2}$. However, as pointed out
before, we will use the fact that these $\gamma $ prefactors will cancel
with factors in the normalizing partition function and directly will only
need to use, in the final expression a lower-rank Stieltjes-Wigert partition
function, as we shall see below.

We need to compute now the Stieltjes transform of these Stieltjes-Wigert
polynomials. Even for classical polynomials, such as Hermite, Laguerre or
Jacobi polynomials, these polynomials are not straightforward to obtain \cite%
{GW,GW2,Fyodorov:2002ub}. In addition, in the Stieltjes-Wigert case, there
are some subtleties associated with giving a fully complete solution for
complex variable $z$\footnote{%
This is related to the undetermined nature of the associated moment problem
and to the fact that, in the Nevanlinna parametrization of the Stieltjes
transform of a measure with log-normal moments, the corresponding Pick
function for the log-normal weight is not explicitly known. See \cite{Christ}
for details.}, but for our matrix model applications here, this will not
play a role.

To obtain the Stieltjes transform of the polynomials we shall use results on
associated Stieltjes-Wigert polynomials. These are defined by%
\begin{equation}
Q_{n}(x):=\int \frac{P_{n}(x)-P_{n}(y)}{x-y}\omega _{\mathrm{SW}}\left(
y\right) dy\text{ \ \ \ }n\geq 0.  \label{SQ}
\end{equation}%
These polynomials have been computed explicitly in \cite{Christ} for a
slightly different definition of the Stieltjes-Wigert polynomials. Adapting
the result there immediately gives%
\begin{equation}
Q_{n}(x)=\frac{(-1)^{n}q^{n/2+1/4}}{\sqrt{(q;q)_{n}}}\sum_{p=0}^{n-1}q^{-%
\frac{p^{2}}{2}}\left( \sum_{j=p+1}^{n}\QATOPD[ ] {n}{j}_{q}(-1)^{j}q^{\frac{%
j^{2}}{2}+\left( p+\frac{1}{2}\right) j}\right) x^{p}.  \label{Q}
\end{equation}%
The difference between these polynomials $Q_{n}(x)$ and the ones we need to
obtain is, manifestly, the Stieltjes transform of the log-normal itself (\ref%
{LN1}), because, it immediately follows from (\ref{SQ}) that%
\begin{equation}
Q_{n}(x)=P_{n}(x)\int \frac{\omega _{\mathrm{SW}}\left( y\right) dy}{x-y}%
-\int \frac{P_{n}(y)\omega _{\mathrm{SW}}\left( y\right) }{x-y}dy.
\label{poly}
\end{equation}%
Thus, we need to compute the first Stieltjes transform in (\ref{poly}). We
proceed by using Mordell's integral \cite{Mordell,Russo:2014bda}, after a
change of variables. Specifically, we need to compute%
\begin{equation}
\overline{I}(\lambda )=\int \frac{\omega _{\mathrm{SW}}\left( x\right) dx}{%
\lambda +x}=\frac{1}{\sqrt{2\pi g}}\frac{e^{m}}{c}\int_{0}^{\infty }\frac{dx%
}{\left( 1+x\frac{e^{m}}{c}\right) }\mathrm{e}^{-\frac{\log ^{2}x}{2g}},
\label{M}
\end{equation}%
with $\lambda \in 
\mathbb{R}
^{+}$ or $\lambda \in 
\mathbb{C}
$. By changing variables $x/c=e^{\mu }$, as in \cite{Russo:2014bda} and
where $c$ is again given by (\ref{CV}), we obtain%
\begin{equation*}
\overline{I}(\lambda (c,m))=\frac{e^{m}e^{-\frac{\log ^{2}c}{2g}}}{\sqrt{%
2\pi g}}\int_{-\infty }^{\infty }d\mu \frac{\mathrm{e}^{\mu \left(
1-N+N_{f}/2+i\eta \right) }}{1+e^{\mu +m}}\mathrm{e}^{-\frac{\mu ^{2}}{2g}},
\end{equation*}%
which is the Mordell integral \cite{Russo:2014bda}. We write it in terms of $%
I(\ell ,m)$ (\ref{imor})%
\begin{equation}
\overline{I}(\lambda (c,m))=\frac{e^{-\frac{g}{2}(N+N_{f}/2-i\eta )^{2}}}{%
\sqrt{2\pi g}}I(\ell =-N-N_{f}/2+i\eta ,m).  \label{mod1mod2}
\end{equation}%
Therefore, gathering all the previous results, and taking into account that
we need the Stieltjes transform of the monic orthogonal polynomials $\pi
_{n}\left( x\right) $ (\ref{monicSW}), we have that%
\begin{equation}
h_{n}(\lambda )=-Q_{n}^{\ast }(\lambda )+\pi _{n}(\lambda )\overline{I}%
\left( \lambda \right) ,  \label{final-h}
\end{equation}%
where $Q_{n}^{\ast }(\lambda )$ is now%
\begin{equation}
Q_{n}^{\ast }(x):=\int \frac{\pi _{n}(x)+\pi _{n}(y)}{x+y}\omega _{\mathrm{SW%
}}\left( y\right) dy\text{ \ \ \ }n\geq 0,
\end{equation}%
which using the result (\ref{Q}) explicitly gives 
\begin{equation*}
Q_{n}^{\ast }(\lambda )=(-1)^{n}q^{-n^{2}-n/2}\sum_{p=0}^{n-1}(-1)^{p}q^{-%
\frac{p^{2}}{2}}\left( \sum_{j=p+1}^{n}\QATOPD[ ] {n}{j}_{q}q^{\frac{j^{2}}{2%
}+\left( p+\frac{1}{2}\right) j}\right) \lambda ^{p},
\end{equation*}%
and hence we have an explicit expression for (\ref{final-h}). Besides, as
shown in detail in \cite{Russo:2014bda}, there are two settings where $%
I(\ell ,m)$ admits a very explicit form in terms of a finite sum expression:

\begin{enumerate}
\item When the $q$-parameter $q=\exp (-g)=\exp (-2\pi i/k)$ is a root of
unity, regardless of $\lambda $, which is the case in the physical setting
of Chern-Simons-matter theory. This is one of the main results by Mordell 
\cite{Mordell}. In this case, and for the finite order $n$ case, the
Stieltjes-Wigert polynomials are still well defined. See below for a
discussion of the large $N$ limit and $q$ roots of unity.

\item For arbitrary $q$ and the mass being $m=pg$ with $p\in 
\mathbb{N}
$ \cite{Russo:2014bda}\footnote{%
We worked in \cite{Russo:2014bda} with $\eta =0$ so we refer here to this
setting.}. In this case, the lambda parameter, which is $\lambda =c\exp
(-m)=\exp (g(N+N_{f}/2-p))$, is an integer or half-integer power of the $q$%
-parameter.
\end{enumerate}

We focus on the first case above, having the following explicit expression
for the Mordell integral for $k>0$ \cite{Russo:2014bda,GM}\footnote{%
There is an analogous expression for $k<0$. Notice also that the prefactor
of $G_{+}$ in \cite{Russo:2014bda} is slightly different. This is due to the
fact that there we discussed the model with $2N_{f}$ hypermultiplets.}%
\begin{equation}
I(\ell ,m)=2\pi \,e^{-i\pi (\ell +\frac{k}{4})}\ e^{-m(\ell +\frac{k}{2})+%
\frac{ikm^{2}}{4\pi }}G_{+}\left( k,1,-\ell -1+i\frac{km}{2\pi }-\frac{k}{2}%
\right) \ ,  \label{mart}
\end{equation}%
with%
\begin{equation}
G_{+}\left( k,1,-\ell -1+i\frac{km}{2\pi }-\frac{k}{2}\right) =\frac{1}{%
e^{-2\pi i\ell -km}-1}\left( -\sqrt{\frac{i}{k}}\ \sum_{r=1}^{k}e^{\frac{%
i\pi }{k}(r-\ell -1-\frac{k}{2}+i\frac{km}{2\pi })^{2}}+i\right) .
\label{mert}
\end{equation}%
We will present the corresponding explicit expression of the polynomial (\ref%
{final-h}) below. The appearance of the Mordell integral is natural and will
also provide consistency checks of the generic formulas obtained here, by
comparison with some of the results in \cite{Russo:2014bda}. Before that, we
show in what follows how to alternatively obtain the polynomials using the
Mordell integral result (\ref{M}) and the three-term recurrence satisfied by
the polynomials.

\subsection{Three-term recurrence and explicit expressions}

There is an alternative way to generate the polynomials $h_{n}(\lambda )$
from the Mordell integral $\overline{I}(\lambda )$. For this, we can use a
fundamental result in the theory of orthogonal polynomials \cite{GW,GW2}:
the polynomials $h_{n}(\lambda )$ satisfy the same three-term recurrence
relation as the Stieltjes-Wigert polynomials%
\begin{equation}
h_{n+1}(\lambda )=(\lambda -b_{n})h_{n}(\lambda )-a_{n}h_{n-1}(\lambda ).
\label{rec}
\end{equation}%
The difference lies on the initial values, which for the $h_{n}(\lambda )$
are%
\begin{equation}
h_{-1}(\lambda )=-1\text{ \ \ and \ }h_{0}(\lambda )=\int \frac{\omega _{%
\mathrm{SW}}\left( x\right) dx}{\lambda +x}=\overline{I}\left( \lambda
\right) ,  \label{Initial}
\end{equation}%
and $a_{0}$ has to be assumed non-zero and equal to $a_{0}=\int \omega
\left( t\right) dt$ \cite{GW,GW2}. The standard Stieltjes-Wigert polynomials
satisfy instead, the customary%
\begin{equation}
S_{-1}(x)=0\text{ \ and \ \ }S_{0}(x)=1.  \label{in2}
\end{equation}%
Hence, the $h_{0}$ polynomial is Mordell's integral $h_{0}(\lambda )=%
\overline{I}(\lambda )$. The use of the three-term recurrence is used for an
algorithmic-computation of the Stieltjes transform of the classical
orthogonal polynomials in \cite{GW,GW2}. Let us use (\ref{rec}) to generate
the first polynomials from Mordell's integral and we shall see that we again
obtain (\ref{final-h}).

The explicit expressions for the coefficients $b_{n}$ and $a_{n}$ of the
monic Stieltjes-Wigert polynomials are%
\begin{eqnarray*}
b_{n} &=&q^{-2n-3/2}(1+q-q^{n+1}), \\
a_{n} &=&q^{-4n}(1-q^{n}).
\end{eqnarray*}%
We can find a few polynomials using the recurrence, leaving for simplicity
the coefficients in generic form%
\begin{equation*}
h_{1}(\lambda )=(\lambda -b_{0})\overline{I}(\lambda )+a_{0},
\end{equation*}%
\begin{equation*}
h_{2}(\lambda )=(\lambda -b_{1})\left[ (\lambda -b_{0})I(\lambda )+a_{0}%
\right] -a_{1}\overline{I}(\lambda )=\overline{I}(\lambda )\left[ (\lambda
-b_{1})(\lambda -b_{0})-a_{1}\right] +(\lambda -b_{1})a_{0},
\end{equation*}%
\begin{equation}
h_{3}(\lambda )=\overline{I}(\lambda )\left[ (\lambda -b_{2})(\lambda
-b_{1})(\lambda -b_{0})-a_{1}(\lambda -b_{2})-a_{2}(\lambda -b_{0})\right]
+a_{0}a_{2}+(\lambda -b_{2})(\lambda -b_{1})a_{0}.  \notag
\end{equation}%
Notice that this construction, based solely on the recurrence, must be
equivalent to the previous one (\ref{final-h}) and hence the polynomial
multiplying the Mordell integral in the examples above must be the
Stieltjes-Wigert polynomial. This is indeed the case since, as noted above,
they satisfy (\ref{rec}) but with (\ref{in2}). Therefore the
Stieltjes-Wigert polynomials follow from these examples above by taking $%
a_{0}=0$ and the Mordell integral to $1$, which proves that the polynomial
multiplying Mordell's integral is the Stieltjes-Wigert polynomial. The rest
is a lower-order polynomial, which corresponds to $Q^{\ast }(\lambda )$
above and is more difficult to obtain from the recurrence, but was easily
characterized above using the results in \cite{Christ}.

Let us now compare the expressions to what is known by the other methods in 
\cite{Russo:2014bda}, which analyzed the case $N_{f}=1$. First we look at
the Abelian cases%
\begin{equation}
\overline{Z}_{N_{f}=1}^{U(1)}=\gamma _{0}h_{0}(\lambda )=q^{1/2}\overline{I}%
(\lambda ).  \label{Ab1}
\end{equation}%
For the matrix model (\ref{Z2}) if $N=N_{f}=1$ then $c=1$ and we have%
\begin{equation}
\widetilde{Z}_{N_{f}=1}^{U(1)}=\frac{-\gamma _{-1}\gamma _{0}}{2\sinh m}%
\left( h_{-1}(\lambda )h_{0}(\mu )-h_{0}(\lambda )h_{-1}(\mu )\right) =\frac{%
q^{1/2}}{2\sinh m}\left( \overline{I}(\lambda )-\overline{I}(\mu )\right) ,
\label{Ab2}
\end{equation}%
where we have used that $\gamma _{-1}=1$ and $\gamma _{0}=q^{1/2}$. The
latter numerical prefactor is what remains of the Stieltjes-Wigert partition
function\footnote{%
Recall that the partition function without the characteristic polynomial
insertion appears as a normalization factor and the Cauchy transform of the
polynomial captures this normalization as well (\ref{h-def}).} when $N=1$ (%
\ref{ZSW}) and hence the results above are as in \cite{Russo:2014bda}.

The most relevant explicit result within our setting is for the $U(N)$
theory given by (\ref{Z}) with $N_{f}=1$, whose characterization in terms of
the polynomial, namely (\ref{h-def}), is now totally explicit%
\begin{eqnarray*}
\frac{Z_{N_{f}=1}^{U(N)}}{A_{N}} &=&-\sum_{p=0}^{N-2}(-1)^{p}q^{-\frac{p^{2}%
}{2}}\left( \sum_{j=p+1}^{N-1}\QATOPD[ ] {N-1}{j}_{q}q^{\frac{j^{2}}{2}%
+\left( p+\frac{1}{2}\right) j}\right) q^{-(N+\frac{1}{2}-i\eta +\frac{ikm}{%
2\pi })p} \\
&&+B_{N}\sum_{j=0}^{N-1}\QATOPD[ ] {N-1}{j}_{q}(-1)^{j}q^{(j-N+i\eta -\frac{%
ikm}{2\pi })j}\left( -\sqrt{\frac{i}{k}}\ \sum_{r=1}^{k}e^{\frac{i\pi }{k}%
(r+N-i\eta -\frac{1}{2}-\frac{k}{2}+i\frac{km}{2\pi })^{2}}+i\right) ,
\end{eqnarray*}%
where we have left the $q$-parameter notation, where\footnote{%
Since $k\in 
\mathbb{Z}
$ one could also use $q=e^{\frac{2\pi i}{k}}$ instead, but notice that the
explicit form of the Mordell integral does depend on the sign of $k$ \cite%
{Russo:2014bda}. For definitiveness, in the text we have given the
expression for $k>0$.} $q=e^{-\frac{2\pi i}{k}}$, in the polynomials and the
normalization constants are%
\begin{eqnarray*}
A_{N} &=&\frac{e^{\frac{i\pi N}{4}}}{k^{N/2}}e^{-\frac{\pi iN}{k}(N+\frac{1}{%
2}-i\eta )(N-\frac{5}{2}+i\eta )}e^{-\frac{3mN}{2}}e^{\frac{\pi i}{k}\frac{%
(N-1)(2N-1)(2N-3)}{3}}\prod\limits_{j=1}^{N-2}(1-e^{-\frac{2\pi i}{k}%
j})^{N-j}, \\
B_{N} &=&\frac{\sqrt{k}e^{-\frac{i\pi }{k}(N+1/2-i\eta )^{2}}e^{-i\pi
(-N+1/4+i\eta +\frac{k}{4})}\ e^{-m(-N-1/2+i\eta +\frac{k}{2})+\frac{ikm^{2}%
}{4\pi }}}{e^{-2\pi i(-N-1/2+i\eta )-km}-1}.
\end{eqnarray*}%
Recall that the overall constant $A_{N}$ comes from the prefactor in the
matrix model representation (\ref{CSM}) and the pure Chern-Simons partition
function (\ref{ZSW}) (see discussion in\ Section 3), whereas the $B_{N}$
accounts for the prefactors, (\ref{mod1mod2}), (\ref{mart}) and (\ref{mert}%
), in the explicit expression of the Mordell integral part of the Cauchy
transform of the Stieltjes-Wigert polynomial (\ref{final-h}).

Precisely, the explicit appearance of the partition function of
supersymmetric pure Chern-Simons theory on $S^{3}$ implies that, as happens
in the case of ABJ theory \cite{Awata:2012jb}, $A_{N}=0$ and, consequently, $%
Z_{N_{f}=1}^{U(N)}=0$ for $N>\left\vert k\right\vert +1$. This is known to
be due to spontaneous supersymmetry breaking and, as pointed out in \cite[%
Remark (4), Section 2.2]{Awata:2012jb}, it will hold for any
Chern-Simons-matter theory that contains supersymmetric pure Chern-Simons
theory as a subsector. Notice that the appearance of the term $\gamma _{N-1}$
in the general solution (\ref{h-def}) implies that $A_{N}$ above contains
the $U(N-1)$, instead of $U(N)$, pure Chern-Simons partition function.

This feature of the orthogonal polynomial result is in agreement with the
explicit Hankel determinant computations in \cite{GM} and can also be
further understood in terms of Giveon-Kutasov duality, because the dual
theory of $Z_{N_{f}=1}^{U(N)}$ for $N>\left\vert k\right\vert +1$ has a
gauge group with negative rank, which signals spontaneous supersymmetry
breaking and implies that $Z_{N_{f}=1}^{U(N)}=0$ in this case \cite[Section
3.3.]{GM}.

The higher flavour cases also have a seemingly simpler expression than those
based on determinants of Mordell integrals and their derivatives \cite{GM}.
For the model (\ref{W}) and $N_{f}=2$ we have%
\begin{eqnarray}
\overline{Z}_{N_{f}=2}^{U(N)} &=&(-1)\gamma _{N-2}\gamma _{N-1}\left\vert 
\begin{array}{cc}
h_{N-1}(\lambda ) & h_{N-2}(\lambda ) \\ 
h_{N-1}^{\prime }(\lambda ) & h_{N-2}^{\prime }(\lambda )%
\end{array}%
\right\vert  \label{2flav} \\
&=&-\gamma _{N-2}\gamma _{N-1}\left( h_{N-1}(\lambda )h_{N-2}^{\prime
}(\lambda )-h_{N-2}(\lambda )h_{N-1}^{\prime }(\lambda )\right) .  \notag
\end{eqnarray}%
We can not evaluate the Abelian case with this expression due to the above
condition $N_{f}\leq N$. Hence, the first simplest particular case of (\ref%
{2flav}) to analyze is $N=2$. Using examples obtained from the recurrence
above, we find, using that $b_{0}=q^{-3/2}$ and $a_{0}=\int \omega
_{SW}\left( t\right) dt=q^{1/4}$ that%
\begin{equation*}
\overline{Z}_{N_{f}=2}^{U(2)}=2(\lambda -q^{-3/2})\overline{I}(\lambda )%
\overline{I}^{\prime }(\lambda )-\overline{I}^{2}(\lambda )+q^{1/4}\overline{%
I}(\lambda ).
\end{equation*}

\subsection{Relationship with the Hankel determinant of Mordell integrals}

The appearance of the Mordell integral in the characterization of the Cauchy
transform of the Stieltjes-Wigert polynomials is natural and consistent, as
we have seen, with the Abelian setting (\ref{Ab1}) and (\ref{Ab2}). In the
non-Abelian case, the partition function has been characterized as a Hankel
determinant of Mordell integrals \cite{Russo:2014bda,GM}. However, in the
solution given before, only one Mordell integral appears, the one in (\ref%
{final-h}).

While a full comparison with \cite{Russo:2014bda,GM} is left for future
work, we can show that indeed, the Hankel matrix in \cite{Russo:2014bda,GM},
while in principle contains $2N$ different Mordell integrals, can actually
be reduced to a matrix with a single Mordell integral. For this, we first
use the modular properties of the Mordell integral and then sum columns and
rows of the matrix. Following the notation in \cite{Mock}, we write (\ref%
{imor}) in the equivalent form, with $z\in 
\mathbb{C}
$ and $\tau \in \mathbb{H}$, then%
\begin{equation}
J(z,\tau )=\int_{%
\mathbb{R}
}dx\exp (\pi \mathrm{i}\tau x^{2}-2\pi zx)/\cosh \pi x,  \label{Mord-Szweg}
\end{equation}%
is the only holomorphic function in $z$ which satisfies \cite{Mock}%
\begin{equation}
J(z)+J(z+1)=\frac{2}{\sqrt{-\mathrm{i}\tau }}e^{\pi \mathrm{i}\left(
z+1/2\right) ^{2}/\tau },  \label{first-mod}
\end{equation}%
\begin{equation*}
J(z)+e^{-2\pi \mathrm{i}z-\pi \mathrm{i}\tau }J(z+\tau )=2e^{-\pi \mathrm{i}%
z-\pi \mathrm{i}\tau /4}.
\end{equation*}%
We consider the matrix model (\ref{Z}) with $N_{f}=1$, with $m=0$ and $\eta
=0$ for clarity of exposition. Taking into account that%
\begin{equation*}
\prod_{i<j}\left( e^{\mu _{i}}-e^{\mu _{j}}\right) =\prod_{i}e^{(N-1)\mu
_{i}/2}\prod_{i<j}2\sinh (\frac{1}{2}(\mu _{i}-\mu _{j})),
\end{equation*}%
and the change of variables $e^{\mu _{i}}=x_{i}$, together with Andreief
identity, which expresses an Hermitian random matrix ensemble in terms of a
Hankel determinant \cite{For,Deift}, we have that%
\begin{equation*}
Z_{N_{f}=1}^{U(N)}(m=\eta =0)=\det \left[ J(z(i,j),\tau )\right]
_{i,j=0}^{N-1},
\end{equation*}%
where $J(z(i,j),\tau )$ is (\ref{Mord-Szweg}) with $z(i,j)=N-1-i-j$ and $%
\tau =k.$ Hence, writing $z$ as $N-1$, the Hankel matrix is of the type 
\begin{equation*}
\begin{pmatrix}
J(z) & J(z-1) & J(z-2) & ... & J(z-N+1) \\ 
J(z-1) & J(z-2) & J(z-3) & ... & J(z-N) \\ 
J(z-2) & J(z-3) & J(z-4) & ... & J(z-N-1) \\ 
\vdots & \vdots & \vdots & \ddots &  \\ 
J(z-N+1) & J(z-N) & J(z-N-1) &  & J(z-2N+2)%
\end{pmatrix}%
.
\end{equation*}%
By using systematically the modular property (\ref{first-mod}) and summing
rows and columns, all the Mordell integrals, except one, can be traded for
sums of the Gaussian term in the r.h.s. of (\ref{first-mod}). For example,
for $N=3$ and defining the notation $G_{n}:=2e^{\pi \mathrm{i}\left(
z-n/2\right) ^{2}/k}/\sqrt{-\mathrm{i}k}$ we have that%
\begin{equation*}
Z_{N_{f}=1}^{U(3)}=\det 
\begin{pmatrix}
J(z) & G_{1} & G_{3} \\ 
G_{1} & G_{3}+G_{1} & G_{3}+G_{5} \\ 
G_{3} & G_{5}+G_{3} & G_{5}+G_{7}%
\end{pmatrix}%
.
\end{equation*}%
This determinant can be further simplified due to the immediate relationship
between the $G_{n}$, but the point to stress here is that it does have the
same form as the Cauchy transform of the Stieltjes-Wigert polynomial. We
will discuss in further detail the relationship between the two methods
elsewhere.

\subsection{Comments on the polynomials at large N}

Already in the original paper of 1923, Wigert proved that the polynomials (%
\ref{SW}) have the limiting behavior \cite{Wigert}%
\begin{equation}
\lim_{n\rightarrow \infty }\left( -1\right) ^{n}q^{-n^{2}-n/2}\pi
_{n}(x;q)=\sum_{k=0}^{\infty }(-1)^{k}\frac{q^{k^{2}+k/2}}{(q;q)_{k}}x^{k},
\label{P-ent}
\end{equation}%
which is essentially the $q$-Airy function \cite{LiWong}%
\begin{equation*}
A_{q}(z)=\sum_{k=0}^{\infty }\frac{q^{k^{2}}}{(q;q)_{k}}(-z)^{k}.
\end{equation*}%
In general, the two polynomials $P_{n}(\lambda )$ and $Q_{n}(\lambda )$
above, which conform together the analytical solution to the matrix model,
converge uniformly on compact subsets of $%
\mathbb{C}
$ when $n\rightarrow \infty .$ In fact, they tend to an entire function when 
$q$ is real as we have seen above for $P_{n}(x;q)$ (\ref{P-ent}) and also,
for $n\rightarrow \infty $%
\begin{equation*}
\lim_{n\rightarrow \infty }\left( -1\right)
^{n}q^{-n^{2}-n/2}Q_{n}(x;q)=\sum_{p=0}^{\infty }(-1)^{p}q^{-\frac{p^{2}}{2}%
}\left( \sum_{j=p+1}^{\infty }\frac{q^{\frac{j^{2}}{2}+\left( p+\frac{1}{2}%
\right) j}}{(q;q)_{j}}\right) x^{p}.
\end{equation*}%
The problem with these expressions for the physical case, where $q$ is a
root of unity, is that while the series are analytic inside $\left\vert
q\right\vert <1$ they have $\left\vert q\right\vert =1$ as a natural
boundary. The more subtle behavior of such $q$-series when $\left\vert
q\right\vert =1$ has only been addressed more recently \cite{DSL}. Notice,
for example, that for $q$ root of unity, the power series coefficients above
are not even defined but, on the other hand, (\ref{P-ent}) contains, when $%
x=q^{-1/2}$ and $x=q^{1/2}$, the two famous series that conform the
Rogers-Ramanujan identities and it is known that these series admit for
example a convergent continued fraction expansion even for $q$ root of unity.

Likewise, the non-trivial asymptotics of the monic Stieltjes-Wigert
polynomials have been worked out explicitly only in recent years (see \cite%
{LiWong} and references therein) and it would be interesting to extend such
computations to the relevant setting here, which involves $q$ root of unity
and requires the asymptotics of the Cauchy transform of the polynomials.

\section{Semiclassical limit}

We have shown so far that the matrix model (\ref{Z}) is the average of a
negative moment of a characteristic polynomial (or more generally, the
correlation function of a characteristic polynomial) in a Stieltjes-Wigert
ensemble \cite{Tierz}. On the other hand, it is well-known that the
Stieltjes-Wigert matrix model is, in the semiclassical limit, a Gaussian
matrix model, also known as GUE ensemble \cite{For}. Thus, we expect the
semiclassical limit of (\ref{Z}) to be a certain average of a characteristic
polynomial on a GUE ensemble.

We now study the semiclassical limit of the Chern-Simons-matter matrix model
(\ref{Z}).\ Such limit can be obtained by firstly performing the change of
variables in the matrix model%
\begin{equation*}
\mu _{i}\rightarrow \sqrt{2g}\mu _{i}
\end{equation*}%
in order to eliminate factors of $g$ from the Gaussian weight function in (%
\ref{Z}), and then expanding for small coupling, which indeed corresponds to 
$k\rightarrow \infty $.

Doing so, we effectively turn the small-$g$ expansion into a small-$\mu $
expansion, affecting only those factors of $\mu $ which were not multiplied
by $1/g$ in the original matrix integral. For simplicity we take the FI
parameter $\eta =0$, but it can be elementary accounted for as well. Taking
all this into account, one can see that the terms to be approximated within
the matrix model are the Vandermonde%
\begin{equation}
\prod_{i<j}4\sinh ^{2}\left( \frac{1}{2}\left( \mu _{i}-\mu _{j}\right)
\right) =\prod_{i<j}\left( \mu _{i}-\mu _{j}\right) ^{2}+\prod_{i<j}\frac{1}{%
12}\left( \mu _{i}-\mu _{j}\right) ^{4}+\text{ }...\,,  \label{eq:coshexp}
\end{equation}%
and the matter part of the matrix model:%
\begin{equation}
\left( 2\cosh \left( \frac{1}{2}\left( \mu _{i}+m\right) \right) \right)
^{-N_{f}}=\left[ \frac{\sinh \left( \frac{m}{2}\right) }{\sinh (m)}\right]
^{N_{f}}\left( 1-\frac{1}{2}\mu _{i}N_{f}\tanh \left( \frac{m}{2}\right)
\right) +\text{ }..\,.  \label{expa-matt}
\end{equation}%
It is worth mentioning that the mass $m$ induces a $\mathcal{O}(\mu )$
contribution in (\ref{eq:coshexp}), which allows us to neglect the second
term on the right-hand side of (\ref{eq:coshexp}) when considering the
lowest possible order. Thus, this result is valid for the massive case $%
m\neq 0$. Notice that in this section we can relax the assumption of $%
\mathcal{N}=3$ supersymmetry and consider also the case $\mathcal{N}=2$ with
the same matter content. The reason is that corrections to the UV R-charges $%
R_{f}=1/2$ appear at $\mathcal{O}\left( \frac{1}{k^{2}}\right) $ \cite%
{Jafferis:2010un,Amariti:2011da}, and thus one can neglect its effect in the
leading large-$k$ limit. Hence, at leading order even the $\mathcal{N}=2$
theory does not get corrections from the RG flow mixing of the $R$-charge
with other Abelian symmetries. Therefore, the resulting matrix model is%
\begin{equation}
Z_{N_{f},k\rightarrow \infty }^{U(N)}=C_{N}^{N_{f}}(m)\int_{-\infty
}^{\infty }d^{N}\mu \prod_{i<j}\left( \mu _{i}-\mu _{j}\right) ^{2}e^{-\frac{%
1}{2g}\sum_{i=1}^{N}\mu _{i}^{2}}\prod_{i=1}^{N}\left( \frac{2}{N_{f}\tanh
\left( \frac{m}{2}\right) }-\mu _{i}\right) \,,  \label{Z-char}
\end{equation}%
where the prefactor is%
\begin{equation*}
C_{N}^{N_{f}}(m)\equiv \frac{(-1)^{N}}{\left( 2\pi \right) ^{N}N!}\left[
\left( 2\cosh (m/2)\right) ^{-N_{f}}\frac{N_{f}}{2}\tanh \left( \frac{m}{2}%
\right) \right] ^{N}.
\end{equation*}%
Hence the result is that%
\begin{equation*}
Z_{N_{f},k\rightarrow \infty }^{U(N)}=C_{N}^{N_{f}}(m)G_{N}\pi _{N-1}^{(H)}(%
\frac{2}{N_{f}\tanh \left( \frac{m}{2}\right) }),
\end{equation*}%
where $\pi _{N-1}^{(H)}\left( x\right) $ denotes the monic Hermite
polynomial of order $\left( N-1\right) $, orthogonal with regards to the
weight function $\omega \left( x\right) =e^{-x^{2}/2g}$ \cite{szego} and $%
G_{N}$ is the Barnes G-function, which is the partition function of the
Gaussian matrix model without the characteristic polynomial insertion \cite%
{For} (GUE ensemble; recall such normalization from\ Section 2). Note that
we recover the pure Chern-Simons partition function in the semiclassical
limit (which is indeed the Barnes function), upon setting $N_{f}=0$ in (\ref%
{Z-char}), as expected.

More specifically, the Hermite polynomials orthogonal w.r.t. to the weight
above are given by%
\begin{equation*}
h_{n}(x)=(-1)^{n}e^{\frac{x^{2}}{2g}}\frac{d^{n}}{dx^{n}}\left( e^{-\frac{%
x^{2}}{2g}}\right) =g^{-n}x^{n}+...
\end{equation*}%
and hence $\pi _{n}^{(H)}\left( x\right) =$ $g^{n}h_{n}(x)$ and the $%
h_{n}(x) $ are related to the Hermite polynomials $H_{n}(x),$ which are the
ones orthogonal w.r.t. to the usual unscaled weight $\omega \left( x\right)
=e^{-x^{2}}$, by $h_{n}(x)=(2g)^{-n/2}H_{n}(x/\sqrt{2g})$. Thus%
\begin{equation}
\pi _{n}^{(H)}\left( x\right) =\left( \frac{g}{2}\right) ^{n/2}H_{n}(x/\sqrt{%
2g})  \label{rescalingH}
\end{equation}%
and the explicit polynomial expression for the partition function is%
\begin{equation*}
\frac{Z_{N_{f},k\rightarrow \infty }^{U(N)}}{G_{N}}=\frac{(-1)^{N}}{\left(
2\pi \right) ^{N}N!\left( 2\cosh (m/2)\right) ^{N_{f}N}}\sum_{j=0}^{\left%
\lfloor \frac{N-1}{2}\right\rfloor }\frac{(-1)^{j}\left( \pi i/k\right) ^{j}%
}{j!(N-1-2j)!}\left( \frac{N_{f}\tanh \left( m/2\right) }{2}\right) ^{2j+1},
\end{equation*}%
where $\left\lfloor x\right\rfloor $ denotes the largest integer not greater
than $x$ (floor function).

\subsection{Chiral models}

An important aspect of this solution in terms of Hermite polynomials is that
the same type of expansion (\ref{expa-matt}) holds for the double sine
functions $s_{b=1}(\sigma )$ in (\ref{Zds}) (see also \cite{Amariti:2011da}
for an expansion of such type). This implies that the semiclassical limit of
the more general model (\ref{Zds}) admits a very similar analysis. These
models are named chiral in a somewhat loose sense \cite{Benini:2011mf}. Let
us explore a few particular cases. In the same limit as above, but for the
matter content in (\ref{Zds}) with $N_{f}\neq \bar{N}_{f}$, and using that $%
s_{b=1}(z)=\exp (l(z))$, where \cite{Jafferis:2010un}%
\begin{equation*}
l(z)=-z\log (1-e^{2\pi iz})+\frac{i}{2}\left( \pi z^{2}+\frac{1}{\pi }%
\mathrm{Li}_{2}(e^{2\pi iz})\right) -\frac{i\pi }{12},
\end{equation*}%
we obtain%
\begin{equation}
\frac{\left( s_{b=1}(\frac{i}{2}-\mu _{i}-m)\right) ^{N_{f}}\left( s_{b=1}(%
\frac{i}{2}+\mu _{i}+m)\right) }{A(m,N_{f},\bar{N}_{f})}^{\bar{N}%
_{f}}=\left( 1-\frac{1}{4\pi }\tanh \left( \frac{m}{2}\right) \left[
im\left( N_{f}-\bar{N}_{f}\right) +\left( N_{f}+\bar{N}_{f}\right) \pi %
\right] \mu _{i}\right) +\text{\ldots }  \label{sexp}
\end{equation}%
where $A(m,N_{f},\bar{N}_{f})$ is a prefactor, easily computed. Thus, the
partition function is also given by a Hermite polynomial, but with a
(complex) spectral parameter: 
\begin{equation}
\lambda =\frac{4\pi }{\tanh \left( \frac{m}{2}\right) \left[ im\left( N_{f}-%
\bar{N}_{f}\right) +\left( N_{f}+\bar{N}_{f}\right) \pi \right] }.
\label{spectral}
\end{equation}%
The expansion (\ref{sexp}) when $\bar{N}_{f}=N_{f}$ leads to the same result
as above (\ref{expa-matt}), as it should, due to the identity (\ref{dsid}).
Indeed, the spectral parameter (\ref{spectral}) reduces to the one in (\ref%
{Z-char}) when $N_{f}=\bar{N}_{f}$ .

If we choose $N_{f}\neq \bar{N}_{f}$ and also two different masses $m_{1}$
and $m_{2}$ we still have the same interpretation in terms of a Hermite
polynomial, since the same type of expansion (\ref{sexp}) holds. The
spectral parameter of the Hermite polynomial in this case is then%
\begin{eqnarray*}
\widetilde{\lambda } &=&\frac{4\pi }{i\left[ m_{1}N_{f}\tanh \left( \frac{%
m_{1}}{2}\right) -m_{2}\bar{N}_{f}\tanh \left( \frac{m_{2}}{2}\right) \right]
+\left( N_{f}\tanh \left( \frac{m_{1}}{2}\right) +\bar{N}_{f}\tanh \left( 
\frac{m_{2}}{2}\right) \right) \pi } \\
&=&\frac{4\pi }{\tanh \left( \frac{m_{1}}{2}\right) N_{f}\left( im_{1}+\pi
\right) +\bar{N}_{f}\tanh \left( \frac{m_{2}}{2}\right) \left( -im_{2}+\pi
\right) },
\end{eqnarray*}%
which reduces to the two previous particular cases in the corresponding
limit. Notice that these semiclassical results also hold for the matrix
model corresponding to the squashed-sphere \cite{Hama:2011ea} because the
biorthogonal Vandermonde there, has the same semiclassical expansion (\ref%
{eq:coshexp}). Notice that an Hermite polynomial of a complex variable $%
H_{n}(z)$ is actually orthogonal with regards to a complex Gaussian weight 
\cite{Ledoux}, but both Heine identity (\ref{averagecharpol:intro}) and the
asymptotics of the polynomials that follows from its consideration, holds
for a complex spectral parameter $z$ (see for example \cite{BI}). Further
work along this line is an interesting open problem.

Note also that the matrix model in (\ref{Z-char}) represents the insertion
of a FZZT D-brane in the simplest minimal string theory, namely the $%
(p,q)=(2,1)$ minimal model, also known as topological gravity \cite%
{Maldacena:2004sn}. Since we have seen that the Hermite polynomial
characterization holds also for the more general matrix model (\ref{Zds}),
including cases with $N_{f}\neq \bar{N}_{f}$ and with different masses for
the fundamental and anti-fundamental multiplets\footnote{%
The same will hold for $R$-charges different than $1/2$, but this may not be
so interesting in the semiclassical limit.} further investigation of this
description would be interesting, specially since the asymptotics of $%
H_{k}(x)$ orthogonal with respect to a weight $\,e^{-Nx^{2}}$ in the limit $%
k,N\rightarrow \infty $ with $k/N\rightarrow c$ and $x$ fixed are very
well-known and can be applied directly to our setting, in contrast to the
polynomials of the previous Section.

\section{Conclusions and Outlook}

Using the method of characteristic polynomials in random matrix ensembles,
we have given exact analytical expressions for matrix models of
Chern-Simons-matter theory, described in the Introduction. The solution is
directly in terms of orthogonal polynomials, which we have also computed
analytically. The appearance of the polynomials themselves as the solution
(and not just their coefficients, as in pure Chern-Simons theory \cite{Tierz}%
) not only leads to exact solutions of the models for finite $N$ and $N_{f}$
but also opens up the possibility of further interpretation of the
observables of the gauge theory, since the orthogonal polynomials satisfy a
number of properties. For example, they are solution of a differential
equation, satisfy a recurrence relation and their large $N$ for both $q$
fixed and for different scaling limits of the parameters can be carried out
explicitly. However, as pointed out above, to do the latter for $q$ root of
unity is an open problem.

One possible further extension/application of the analytical results
obtained here could consist in applying the same characteristic polynomial
interpretation/method to the case of ABJ and ABJM matrix models. While
already a number of detailed exact solutions for finite $N$ exist for these
models \cite{Awata:2012jb} (see \cite{Hatsuda:2015gca} for a review) this
method has the potential to offer complementary analytical results. Notice
the mathematical similarity between the models (\ref{Z}) and (\ref{Zm2}) and
the ABJ(M) matrix models. The latter are two-matrix models but one could
integrate over one set of the eigenvalues in the ABJ or ABJM matrix model,
using either the Mordell integral method \cite{Russo:2014bda,GM} or using
the orthogonal polynomial results here. In such a first integration, the
other set of eigenvalues would be interpreted as generic masses and then an
additional integration over such masses would complete the calculation.

In addition, given the interest of the ABJ matrix model in the so-called
high-spin limit, due to the conjectured existence of a gravity dual \cite%
{Hatsuda:2015lpa}, it would be worth to see if the semiclassical behavior
explored here for our simpler models, can also be put to use in that
context. Such a putative application to ABJ(M) matrix models requires a
detailed comparison between the results here and the results in \cite%
{Russo:2014bda,GM}. That is, an extension of Section 4.2.

Finally, we have just pointed out the most straightforward interpretation of
the characteristic polynomial point of view of the matrix models in the
context of topological strings as amplitudes of non-compact B-branes \cite%
{Hyun:2006dr} and, in the semiclassical setting of last Section, as FZZT
branes in topological gravity \cite{Maldacena:2004sn}. It would be
interesting if an additional physical picture can be extracted from the
analytical results.


\subsection*{Acknowledgements}


The author is indebted to Luis Melgar and Georgios Giasemidis for valuable
collaboration and for a careful reading of the manuscript. Thanks to Jacob
S. Christiansen for mathematical explanations and a copy of \cite{Wigert}
and to Jorge Russo for discussions and comments. This work was partially
supported by FCT-Portugal through its program Investigador FCT IF2014, under
contract IF/01767/2014. We also gratefully acknowledge support from the
Simons Center for Geometry and Physics, Stony Brook University and from the
Institute for Advanced Study, Princeton, at which some of the research for
this paper was performed. Thanks to Jac Verbaarschot and Mauricio Romo,
respectively, for warm hospitality at these institutions.


\end{document}